\documentclass[10pt]{article}

\usepackage{amsmath}
\usepackage{amssymb}

\usepackage{graphicx}

\usepackage{cite}

\usepackage{color} 

\usepackage{color}
\newcommand{\rev}[1]{{\textcolor{black}{#1}}}

\usepackage[labelfont=bf,labelsep=period,justification=raggedright]{caption}

\makeatletter
\renewcommand{\@biblabel}[1]{\quad#1.}
\makeatother

\date{}

\begin{document}

\begin{flushleft}
{\Large
\textbf{Directedness of information flow in mobile phone communication networks}
}
\\
Fernando Peruani$^{1}$, 
Lionel Tabourier$^{2}$ 
\\
\bf{1} Max Planck Institute for the Physics of Complex Systems, Dresden, Germany
\\
\bf{2} LIP6, Universit\'e Pierre et Marie Curie, Paris, France
\\
\end{flushleft}

\section*{Abstract}

Without having direct access to the information that is being exchanged, traces of information flow can be obtained 
by looking at temporal sequences  of user interactions. These sequences can be represented as causality trees 
whose statistics  result from a complex interplay between the topology of the underlying (social) network and the time correlations among the communications. 
Here, we study causality trees in mobile-phone data, which can be represented as a dynamical directed network. 
This representation of the data reveals the existence of super-spreaders and super-receivers. 
We show that the tree statistics, respectively the information spreading process, are extremely sensitive to the in-out degree correlation exhibited by the users.
We also learn that a given information, e.g., a rumor, would require users to 
 retransmit it for more than 30 hours in order to 
cover a macroscopic fraction of the system. 
Our analysis indicates that topological node-node
correlations of the underlying social network, 
while allowing the existence of information loops, they also promote
information spreading.
%
%
Temporal correlations, and therefore causality effects, are only visible as local phenomena and during short time scales.
These results are obtained through a combination of theory and data
analysis techniques.


\section*{Introduction}
Phone call activity patterns are a manifestation of our complex social dynamics. 
Several aspects of our social behavior are reflected in these communication
patterns, like day-night cycles, high activity at the end of working hours,
or even our mobility
patterns~\cite{onnela2007structure,onnela2007analysis,gonzalez2008understanding,song2010limits,karsai2010small,
lambiotte2008geographical,candia2008uncovering,miritello2010dynamical,bagrow2011collective}. %
Mobile phone data provides an excellent ground to study several interesting
social processes such as, for instance, the spreading of news and rumors, which is the focus of this work.
We start out by asking ourselves whether such phenomenon really occurs through the mobile phone network. 
A phone call certainly involves information exchange between two individuals, but is there information propagation involving more than a single phone call? 
Is it possible to answer this question without having access to the content of
conversations?

Mobile phone log data consists in who calls whom and when, see 
Fig.~\ref{fig:sketch}A. 
A natural way of representing this data is through the use of directed edges. 
For example, let us use $i \to j$ to represent that 
 user $i$ has called user $j$. 
In addition, we have to associate to each directed edge a time series that symbolizes when (and how many times) 
this action took place.
This procedure provides us with a representation of log data in terms of a directed network. 
%


Due to privacy issues we cannot know which information 
is exchanged during phone calls. 
This constraint forces us to adopt a hypothesis regarding how information flows on the network. 
It has been argued that depending on the nature of the information, its propagation dynamics is different~\cite{castellano2009statistical}. 
For example, a political opinion, a fad, a rumor, or a gossip, are supposed to involve,  each of them,  
a different kind of human interaction dynamics which results in a different and particular propagation mechanism~\cite{castellano2009statistical}. 
Here, we assume that the information that is exchanged is either a rumor or news.

The spreading of rumors and news is believed to resemble the spreading of an infectious disease~\cite{anderson1991infectious},  \rev{
which implies the existence of a ``pass-along'' dynamics, as the one observed in email chain letters}.
Rumor spreading models~\cite{daley1964epidemics,liu2003propagation,moreno2004dynamics,moreno2004efficiency} assume that there are two categories of users, those who are informed and those who are not. 
Among informed users, there are in turn two sub-categories: users that are actively broadcasting the rumor, and users that become inactive. 
Several mechanisms for switching from active to inactive spreading behavior have been proposed~\cite{castellano2009statistical, daley1964epidemics,liu2003propagation,moreno2004dynamics,moreno2004efficiency}. 
Given the lack of empirical evidence to support a particular switching mechanism, here we adopt the simplest possible assumption already 
proposed for infectious disease~\cite{anderson1991infectious}: there is a characteristic time $\tau$ after which an active spreader turns into inactive. 
There is, however, a deeper reason to use such a switching mechanism. 
According to this description, we can represent the sequence of events that propagates the infection as a \textit{causality tree}.
Our goal is to study causality trees as a proxy to understand information spreading, and for this we need 
a characteristic time scale which can be easily controlled. 
The parameter $\tau$, which we refer to as \textit{monitoring time}, serves this purpose. 

The mobile phone data, particularly the existence of directed edges, poses the question whether during a phone call the information exchange exhibits a favored flow direction. 
Clearly, the information can flow from the caller to the callee, from the callee to the caller, or in both direction. 
If we think of a rumor being spread on the mobile phone network, and a phone call that involves an informed and an ignorant user, we can imagine that after the communication, both users are informed. 
This picture implies that non intentional spreading of the rumor can occur: 
if the caller is the ignorant user and get the information from the
callee, then the phone call was not intended to propagate the rumor. 
This would mean that rumor spreading does not involve causality and it occurs
without ``intentionality''. 
A different scenario is one in which the spreading is exclusively
active and intentional, and phone calls made by informed individuals are intended to propagate
the information. 
In this scenario, information propagates through active broadcasting, and
involves causality.  
A user can get informed exclusively if he/she is called by an informed
individual. 
Thus, in this framework, causality trees describe information flow. 

At this point we would like to mention two very recent works on information
spreading on mobile phone networks that are particularly
related to our study~\cite{karsai2010small,miritello2010dynamical}. 
Karsai et al.~\cite{karsai2010small} studied flooding of information in a mobile phone network. They assumed 
that users retransmit constantly the information they receive, and estimated
the time required to inform everybody in the system. 
They concluded that the presence of community structures, i.e., topological
correlations, and bursty phone call activity slow down the
spreading of information. 
In~\cite{miritello2010dynamical}, Miritello et al. studied the propagation of information
that obeys a Susceptible-Infected-Removed (SIR) epidemic dynamics. 
They assumed that during a phone call information flows in both direction 
and made use of Newman's theory for disease spreading on undirected complex 
networks~\cite{newman2002spread}  to interpret their results~\footnote{For computational
  purposes, they analyzed the data using directed edges for information
  transmission, but assuming that transmission in directed and undirected
  networks is comparable.}. 
They exclusively focused on the average size of the outbreaks, and confirmed
the results obtained in~\cite{karsai2010small}. 
%

\rev{Here, our main goal is to explore the possibility of intentional propagation of information with a ``pass-along'' dynamics, e.g., a rumor. 
This intentionality implies that the information flow is expected to be given by the directed edges.   
Nevertheless, for completeness and comparison purposes we also consider the possibility that information flows in both
directions along edges. 
We focus on the topological properties of the causality trees resulting from the above described dynamics. 
%
We point out that we aim at characterizing the spreading properties of the system, 
but not at identifying particular events. 
Such a task would require the exploration of the mobile phone data both, forward and backward in time, as explained in~\cite{kovanen2011}. 
We find that the existence of super-spreaders and super-receivers makes the spreading process extremely sensitive to the 
in-out degree correlation of the users. We make use of a simple tree theory to prove this fact.  
We also observe that time correlations do not play a significant role on the statistics of sizes and depths of trees. }
On the other hand, we observe that at short time-scales, i.e., small $\tau$ values, 
the spreading dynamics, respectively the tree statistics, is not sensitive to
topological node-node correlations and can be described simply in terms of the out degree
distribution of the underlying social network. 
Only at large time scales these node-node correlations become dominant,
enhancing the  spreading of information and allowing the circulation of
information in (closed) loops.  
Time-correlations, while they do not have a significant impact on information
spreading, promote the existence of local  information loops. 
It is only at this level that we observe genuine causality effects.


\section*{Results}

\subsection*{Causality trees}
\begin{figure}[!ht]
\centering\resizebox{8.5cm}{!}{\rotatebox{0}{\includegraphics{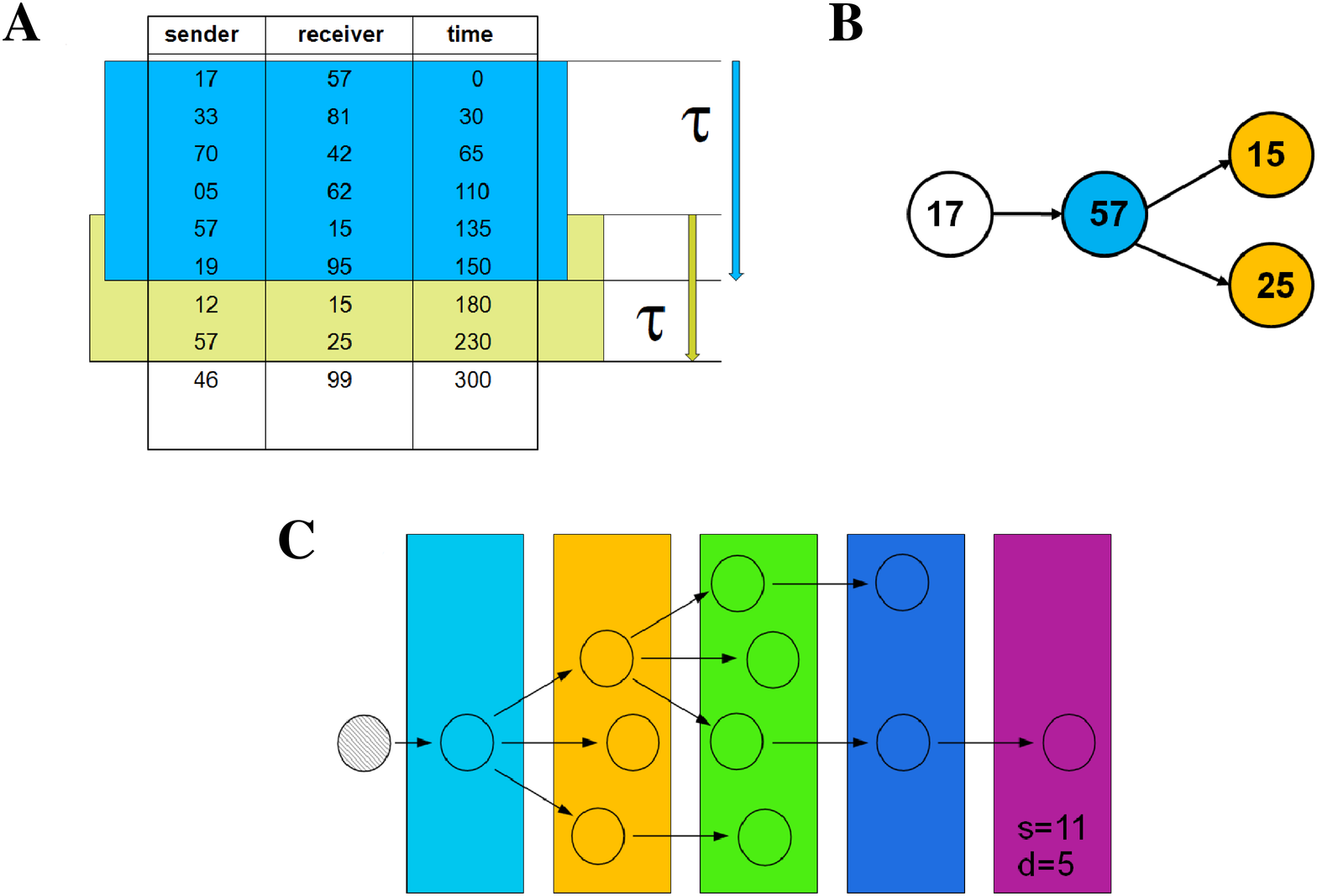}}}
\caption{{\bf Mobile phone data and causality trees.} The mobile phone data, {\bf A}, is processed for a given time-scale  $\tau$ (here $\tau$ = 150) 
  to obtain causality trees, {\bf B}. An example of a real tree of size $s=11$ and
  depth $d=5$ is shown in {\bf C}.} \label{fig:sketch}
\end{figure}

For a given $\tau$, we build the ``causality'' trees, which we
also refer to as  cascades, in the following way. 
We pick up at random a phone call from the database and monitor the activity
of the receiver - e.g.  user {\bf a} - for a period $\tau$. 
%
We register all phone calls  user {\bf a}  makes during this period, and monitor the activity of all users 
who have been called by user {\bf a} during a period $\tau$. 
We repeat this process for every {\it new} user until the cascade gets extinguished. This occurs when
all users in the tree have exceeded the monitoring time $\tau$ and there is no new user to
monitor.   
Figs.~\ref{fig:sketch}A and~\ref{fig:sketch}B illustrate this procedure, which has been
also applied in~\cite{miritello2010dynamical}. 
%
%
Notice that the proposed method is equivalent to inoculate a 
Susceptible-Infected-Recovered (SIR) disease to user {\bf a} and wait for the
infection outbreak to get extinguished. 
This dynamics is similar to the one proposed for rumor spreading as defined in~\cite{moreno2004dynamics}.
Susceptible (i.e., uninformed) users only get infected by
receiving a phone call from an already infected user, i.e.,  phone calls
imply directed links. 
Finally, the transition from the infected to recovered state occurs after a time $\tau$.

%
We focus on two features of the trees: their size $s$ and their depth $d$.  
%
%
The tree size $s$ is simply the number of users forming the tree. 
We use the term {\it depth} to refer to the distance (in terms of nodes) to
the initial node, with  $d$  defined as the maximum depth of the tree, see Fig.~\ref{fig:sketch}C. 

\subsection*{A simple transmission theory}
\subsubsection*{Gaining intuition - a basic  mean-field }
To gain insight into the statistics of the causality trees, we first propose a simple transmission theory relying on the following assumptions.
a) There exists an underlying social network which is static.  
Though we know that at large time scales, e.g., years, the underlying social network is necessarily
dynamic, we assume that at short time scales of the order of hours to few days the
static approximation provides a reasonable description. 
%
%
%
b) Given a couple of nodes $i$ and $j$, there is a directed link from $i$ to $j$ if $i$ called $j$ at least once in the database, i.e, if the directed link is present in the underlying social network. 
This directed link is associated to a time series: the timestamps at which $i$ has contacted $j$. 
For every $i \to j$ link we can define a communication rate $\rho_{i \to j}$ that
indicates the rate at which the link is active to transmit information. 
This is simply the number of phone calls  $i \to j$ that occurred in the database, 
divided by the total time $t_{\infty}$.
c) We define causality trees as sequences of consecutive phone calls where the time
difference between consecutive phone calls is always less or equal to our
observation time $\tau$.

\rev{ 
In this introductory subsection, we aim at gaining 
 some intuition on the spreading process 
by reducing  all the complexity of the social network topology to its 
average degree, and all the complexity associated to the temporal phone call dynamics to the average phone
call rate per edge. 
We focus initially only on the average tree size. 
In the next subsection we will include more complex features such as 
in- and out-degree distributions, in-out degree correlations, etc, and focus on 
more interesting measurements as the size and depth distributions of the trees. 
For the moment, nevertheless, we study this oversimplified dynamics since 
it allows us  to understand in very simple terms that there is a
competition between the monitoring time $\tau$ and the phone call frequency. 
This competition determines the critical monitoring time $\tau_c$ above which a phase transition occurs.
}
The mean-field associated with this simplified process reads:
\begin{eqnarray}\label{eq:mfa}
\begin{array}{ccc}
\dot{S} = - \langle \rho \rangle  \langle k_o \rangle_{\infty} S I\, , & 
\dot{I} = \langle \rho \rangle  \langle k_o \rangle_{\infty} S I  - \frac{I}{\tau} \, , &
\dot{R} =  \frac{I}{\tau} \, ,
\end{array}
\end{eqnarray}
where the dots indicate time derivatives, $S$ refers to the fraction of
individuals that  are uninformed at time $t$, 
$I$ to those that are informed and retransmitting 
 the information, and $R$ to those that got the information and stop 
 retransmitting, $ \langle \rho \rangle$ is the 
average $\rho_{i \to j}$, and $\langle k_o \rangle_{\infty}$ the average (out-)degree. 
Notice that for directed networks, the average in- and out-degree are the same. 

The average cascade size $\langle s \rangle$ is the average number of
individuals that got the information once the process is extinguished
, i.e., when there is no more users retransmitting the information and $I=0$.  
%
%
%
It is easy to realize from Eq.~(\ref{eq:mfa}) that $S(t)=\exp
\left[-\langle \rho \rangle \langle k_o \rangle_{\infty} \int_{0}^{t}I(t')dt' \right]$, 
and $\tau\,R(t)=\int_0^{t} I(t') dt'$. 
Using the fact that $S+I+R=1$ and $I(t\to\infty)=0$, we obtain that the number
of users that once got the information, i.e., the average tree size, $\langle s \rangle = N
R(t\to\infty)$, reads:
\begin{eqnarray}\label{eq:R}
\langle s \rangle/N = 1 - \exp{\left[- \langle \rho \rangle  \langle k_o \rangle_{\infty} \tau \langle s \rangle/N \right]} \, ,
\end{eqnarray}
where $N$ is the number of users in the system. 
Eq.~(\ref{eq:R}) defines a self-consistency equation for $\langle s \rangle$. 
From this expression we can derive the critical monitoring time $\tau_c$
required to observe infinite tree sizes: 
\begin{eqnarray}\label{eq:mfathreshold}
\tau_c =  \frac{1}{\langle \rho \rangle
\langle k_o \rangle_{\infty}}\,.
\end{eqnarray}
This means that if nodes retransmit the information that they get for a period
$\tau>\tau_c$ the resulting trees can be arbitrarily large. 
It is important to make the distinction between the duration of cascade events, i.e., the time elapsed between the first and last phone call of the tree, and the monitoring time $\tau$. 
For instance, when $\tau\sim\tau_c$ only a small fraction of the cascades percolates and consequently the average tree duration is shorter than $\tau_c$. 
Notice that  the monitoring time $\tau$ refers (and controls)  the individual 
behavior of users as callers. 
In fact, we are asking ourselves how the individual behavior of users  should be in
order to allow a rumor to take over a macroscopic fraction of the system. 
We will come back to this point later on and look for an interpretation of
this relevant quantity. 

Now let us consider the other hypothesis mentioned in the introduction, i.e., let us imagine   
 that information travels in both directions of the directed edges. In this case, 
we can still use Eq.~(\ref{eq:R}) to describe the spreading process, but
parameters have a different meaning. 
The average rate activity $\langle \tilde{\rho} \rangle$ now represents the activity
of an undirected edge, i.e., it is the average of $\tilde{\rho}_{i,j}=\rho_{i\to
  j}+\rho_{j\to i}$. 
Finally, the average out-degree $\langle k_o \rangle_{\infty}$  has to be replaced by the average (undirected) degree $\langle k \rangle_{\infty}$,
since now edges are undirected.
Thus Eq.~(\ref{eq:mfathreshold}) becomes:
$ \tau_c =  1 / ( \langle \tilde{\rho} \rangle (\langle k \rangle_{\infty} -1 ))$.

\subsubsection*{Taking into account topological features - in-out degree
  distributions}

\rev{In the following, we move a step further and take into account the in- and out-degree distributions,  
as well as the in-out degree correlations of the underlying static social network, 
 and the time distribution of phone calls. } 
Our goal is to obtain  an expression for the probability $p(k_i, k_o; \tau)$ of finding a user 
with in-degree $k_i$ and out-degree $k_o$ for a given $\tau$. 
The in-degree $k_i$ represents the number of different users that have called
the user during $\tau$. 
Similarly,  $k_o$ denotes the number of different users that the user has
called during the monitoring time $\tau$. 
This means that we are considering that there is an underlying static network
whose directed links are switched on and off --- 
this dynamics defines a (directed) dynamical network. 

In order to compute $p(k_i,k_o;\tau)$, we need to know the probability that an
edge is activated within a period $\tau$. 
More precisely, we need to know for a node that has
in-degree $k_i'$ and out-degree $k_o'$, the probability per in-edge $T_i(k_i',
\tau)$ and per out-edge $T_o(k_o', \tau)$ of being used within a period $\tau$. 
Then, if for instance $k_i'=3$, the probability that two of the three in-edges are
used while one is not, is  $\left( \begin{array}{c} 3 \\ 2 \end{array}
\right) T_i(3, \tau)^{2} (1-T_i(3, \tau))$. 

We denote the probability of finding a node with in-degree $k_i$ and
out-degree $k_o$: $p_{\infty}(k_i, k_o)=p(k_i, k_o; \tau \to \infty)$. 
The probability $p_{\infty}(k_i, k_o)$ refers to the static underlying network that contains all connections that have occurred  
in the whole database. 
Thus, assuming that the in-degree of a node is uncorrelated with the
out-degree of other nodes, we can express 
 $p(k_i, k_o; \tau)$ as:

\begin{eqnarray}\label{eq:pkiko} 
p(k_i,k_o;\tau) &=& \sum_{k_{i}'=k_i; k_{o}'=k_o}^{\infty} p_{\infty}(k_i', k_o') \\
\nonumber
&& \times \left( \begin{array}{c} k_i' \\ k_i \end{array} \right) T_i(k_i', \tau)^{k_i} (1-T_i(k_i', \tau))^{k_i'-k_i} \\
\nonumber
&&\times \left( \begin{array}{c} k_o' \\ k_o \end{array} \right) T_o(k_o', \tau)^{k_o} (1-T_o(k_o', \tau))^{k_o'-k_o} \, ,
\end{eqnarray}
where we have also assumed that the activity of an edge is independent of the
activity of the other edges, and used the binomial distribution approximation
explained above.

To simplify  Eq.~(\ref{eq:pkiko}), we make another assumption:
the edge activity is independent of the in- or out-degree
of the node, i.e. $T_i(k, \tau)=T_o(k, \tau)=T(\tau)$. 
Edges exhibit a heterogeneous distribution $p(\rho)$ of communication rates $\rho$. 
Let us recall that $\rho$ is the rate at which an edge is used within $ \tau$, i.e., the number
of phone calls through the edge divided the total time $t_{\infty}$. 
We need an estimate for the probability $T(\tau)$ that the edge is used during
$\tau$. 
Knowing $\rho$ and assuming a Poissonian process, the probability that the
edge is used is $1-\exp(-\rho \tau)$~\cite{newman2002spread}. 
\rev{Though this assumption is a strong simplification of the actual process at the level of edge usage, 
it does not imply that we are assuming that users exhibit a Poissonian phone call activity. 
Moreover, since we consider the activity of a user proportional to its out-degree, 
if the out-degree distribution is heavy-tail distributed, so is the user activity distribution.}   
Under these assumptions $T(\tau)$ can be estimated as:
\begin{eqnarray}\label{eq:T} 
T(\tau)=\int \, d\rho \, p(\rho) \, \left(1 - e^{-\rho
    \tau} \right) \, .
\end{eqnarray}
%
%

If now we consider that information travels in both direction of the edges, 
we need to take into account the undirected degree distribution exhibited by the nodes. 
Along similar lines, we can express the probability $p(k;\tau)$ of finding a user of undirected degree $k$ for a given $\tau$ as:
\begin{eqnarray} \label{eq:pkuncorrelated}
p(k;\tau) &=& \sum_{k'=k}^{\infty} p_{\infty}(k') \\
\nonumber
&& \times \left( \begin{array}{c} k' \\ k \end{array} \right) \tilde{T}(k', \tau)^{k}
(1-\tilde{T}(k', \tau))^{k'-k} \, ,
\end{eqnarray}
where $p_{\infty}(k')$ refers to the degree distribution of the undirected
static social network, and $\tilde{T}(k', \tau)$ is the probability that an edge
connected to a node of degree $k'$ is used  during $\tau$.

\subsubsection*{Critical monitoring time $\tau_c$}
We now look for an expression of the critical monitoring time $\tau_c$, which is sensitive to the topological
structure of the underlying static network. 
As before, we start by assuming that the underlying static network is directed
--- below we address the alternative case where information travels in both direction on edges. 
To derive the percolation threshold from Eq.~(\ref{eq:pkiko}), we look for the associated generating function 
%
%
 $g(x,y;\tau)=\sum_j \sum_k x^j y^k p(j,k;\tau)$. 
After exchanging the order of the sums in order to use the binomial expansion, we obtain:
\begin{eqnarray}\nonumber
g(x,y,\tau)&=& \sum_{k_i', k_o'} p_{\infty}(k_i', k_o') \left(1+(x-1)T(\tau)\right)^{k_i'}\\
\label{eq:generating}
&&\times   \left(1+(y-1)T(\tau)\right)^{k_o'} \, .
\end{eqnarray}

The process described here corresponds to a situation where, for a given $\tau$, some edges are 
activated while others remain silent.
We want to know whether the (directed) network of activated edges contains giant trees. 
As  explained in~\cite{schwartz2002percolation}, the condition for having  infinite clusters in (static) directed networks
is  $\langle k_i \, k_o \rangle / \langle k_o \rangle \geq 1$. 
We can evaluate this condition for our dynamical network using
Eq.(\ref{eq:generating}), and recalling that  
  $\langle k_i k_o \rangle(\tau) =
xy\partial^2 g/\partial xy |_{x,y=1}$, and $\langle k_i \rangle(\tau)=\langle
k_o \rangle(\tau) = x\partial^2 g/\partial x |_{x,y=1}$. 
The evaluation of the above mentioned condition leads to: 
\begin{eqnarray}\label{eq:threshold_T}
T(\tau) \geq \frac{\langle k_o \rangle_{\infty}}{\langle k_i k_o
  \rangle_{\infty}} \, ,
\end{eqnarray}
where $\langle k_i k_o \rangle_{\infty} = \sum_{k_i,k_o} k_i k_o
p_{\infty}(k_i,k_o) $ and $\langle k_o \rangle_{\infty} =  \sum_{k_o} k_o
p_{\infty}(k_o)$. 
Notice that if the underlying network does not exhibit a giant
component for $t\to\infty$, condition
 (\ref{eq:threshold_T}) cannot be fulfilled for any $\tau$.  

To gain some intuition, let us assume that $\rho \tau_c << 1$, so that from
Eq.~(\ref{eq:T}) we can approximate $T(\tau)$ as 
$T(\tau)\sim \langle \rho \rangle \tau$ so that:
\begin{eqnarray}\label{eq:thresholdcorrelation}
 \tau_c = \frac{\langle k_o \rangle_{\infty}}{\langle \rho \rangle
 \langle k_i k_o \rangle_{\infty}} \,.
\end{eqnarray}

Let us now consider the following two extreme cases:
a) a fully in-out degree correlated underlying static network, where $k_i = k_o$, for which we
get 
\begin{eqnarray}\label{eq:thresholdfullycorrelated}
\tau_c = \frac{\langle k_o \rangle_{\infty}}{  \langle \rho \rangle
  \langle k^2_o \rangle_{\infty} }\,,
\end{eqnarray}    
and b) a fully in-out degree uncorrelated underlying static network, i.e.,  $p_{\infty}(k_i, k_o) =
p_{\infty}(k_i) p_{\infty}(k_o)$, where we find that   
\begin{eqnarray}\label{eq:thresholdfullyuncorrelated}
\tau_c = \frac{1}{
  \langle \rho \rangle \langle k_o \rangle_{\infty} }\,, 
\end{eqnarray}
which is exactly the mean-field prediction given by Eq.~(\ref{eq:mfathreshold}).

\begin{figure}[!ht]
\centering\resizebox{8 cm}{!}{\rotatebox{0}{\includegraphics{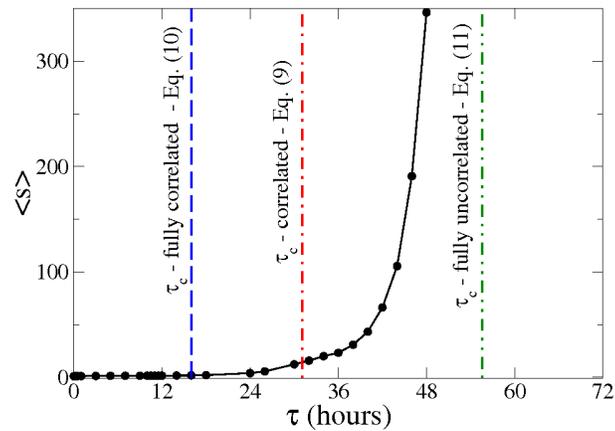}}}
\caption{{\bf Average cascade size as function of the observation time $\tau$.} The
  transition towards arbitrarily large cascade sizes can be estimated
  theoretically using Eq.~(\ref{eq:threshold_T}). The  vertical lines  correspond to
  various assumptions on the underlying social network, particularly on the in-out degree correlation matrix: fully
  correlated, fully uncorrelated, and the actual situation, see text.} \label{fig:firstmoments}
\end{figure}
\begin{figure}[!ht]
\centering\resizebox{11 cm}{!}{\rotatebox{0}{\includegraphics{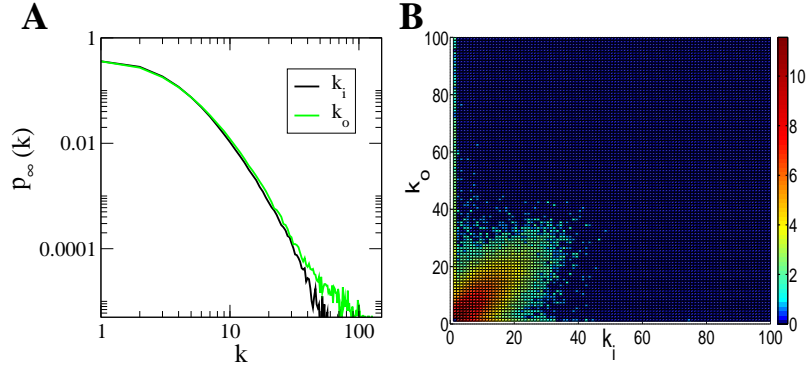}}}
\caption{{\bf Statistics on the in- and out-degree of the underlying social static network.} 
The in- and out-degree distribution, $p_{\infty}(k_i)$ and $p_{\infty}(k_o)$, respectively, are shown in  {\bf A}, while in-out degree correlations, i.e., $p_{\infty}(k_i,k_o)$, 
in  {\bf B},  where the color scale corresponds to  $ln(p_{\infty}(k_i,k_o))$. 
} \label{fig:degree_correlations}
\end{figure}

In Fig.~\ref{fig:firstmoments} we compare the average tree size obtained from the
data with the above discussed theoretical arguments. \rev{The vertical lines correspond to 
the predictions given by Eqs.~(\ref{eq:thresholdcorrelation}), (\ref{eq:thresholdfullycorrelated}), and (\ref{eq:thresholdfullyuncorrelated}).
In order to evaluate Eq.~(\ref{eq:thresholdcorrelation}), we have measured  $\langle k_o \rangle_{\infty}$ and  
$\langle k_i k_o \rangle_{\infty}$ using the underlying static network, and estimated the average edge activity $\langle \rho \rangle$ from the log data. 
For Eq.~(\ref{eq:thresholdfullycorrelated}), we have computed in addition $\langle k_o^2 \rangle_{\infty}$. 
Finally, for Eq.~(\ref{eq:thresholdfullyuncorrelated}), since we assume uncorrelated in-out degree, we only need to know 
 $\langle k_o \rangle_{\infty}$ which has been already measured to estimate Eq.~(\ref{eq:thresholdcorrelation})
 (let us recall that directed networks are such that  $\langle k_o \rangle = \langle k_i \rangle$).
The in- and out-degree distributions of the underlying static network are shown  in Fig.~\ref{fig:degree_correlations}A,
 while the in-out correlation matrix, which is characterized by a  Pearson's coefficient~\cite{kendall1967advanced} of $\sim 0.58$, is shown in Fig.~\ref{fig:degree_correlations}B.}  
Fig.~\ref{fig:firstmoments} shows that there is large difference between the thresholds corresponding to the extreme
cases given by Eqs. (\ref{eq:thresholdcorrelation}) and (\ref{eq:thresholdfullycorrelated}),  more than $35$ hours, 
which indicates that the spreading process strongly depends on the in-out degree
correlations.    
How can we understand that correlations have such an important effect on
the spreading process? 
If we look at either the in- or the out-degree distribution of the underlying static
network, we observe that both of these distributions exhibit fat tails, see Fig.~\ref{fig:degree_correlations}A. 
%
%
The heterogeneity of user degree implies the presence of super-spreaders as
well as super-receivers. 
Though the relevance of super-spreaders have been well identified and
understood since many years~\cite{pastor2001epidemic}, the existence and role of
super-receivers have remained relatively unexplored, 
except for a few noticeable works where the relevance of in-out degree correlations were
acknowledged~\cite{schwartz2002percolation,boguna2005generalized,zamora2008reciprocity}. 
%
%
\rev{ We notice that for an undirected network, the degree of a node plays both roles, being its
in- as well as its out-degree (and consequently, a high degree node 
is both a super-spreader and a super-receiver). 
On the contrary, for a directed network, the difference between the in- and out-degree of a node, as we will see, 
is key to understand the role of the node on the spreading process. 
A super-receiver is a node that is highly susceptible to get the information, 
since it can be contacted by a large number of different users. 
On the other hand, a super-spreader is a node that once informed can
potentially retransmit the information to many different users. 
If a node is a super-receiver but not a super-spreader, it can easily get
the information but it cannot contribute much in the spreading process. 
On the other hand, a node that is a super-spreader but not a super-receiver, 
while it can potentially retransmit the information to many different users, 
it will rarely get the information. 
Consequently, these nodes rarely help to the spreading process of the information. 
In summary, only those nodes that are both, super-spreader and super-receiver, play a significantly role on the spreading process.} 

\rev{For fully in-out correlated networks (see Eq.~(\ref{eq:thresholdfullycorrelated})), nodes exhibit the same in-
and out-degree. 
This implies that all super-receivers are also super-spreaders. 
Thus, each super-spreader has a high probability of getting the information and subsequently retransmit it.
Fig.~\ref{fig:firstmoments}  shows that for $k_i=k_o$ -- Eq.~(\ref{eq:thresholdfullycorrelated}) --
 we would observe arbitrarily large trees for $\tau$ values
larger than $12$ hours. 
In general, we expect larger trees for a fully correlated network 
with the same number of super-spreaders (and
super-receivers) as in the original network. 
Positive in-out degree correlations clearly facilitates
information spreading. 
On the other hand, we expect fully in-out uncorrelated networks 
to exhibit smaller trees, since it is very difficult to find a node that is both, super-spreader as well
as super-receiver. 
Fig.~\ref{fig:firstmoments} indicates that for fully in-out uncorrelated networks -- Eq.~(\ref{eq:thresholdfullyuncorrelated}) -- 
 arbitrarily large
trees would only emerge for $\tau$ values larger than $55$ hours, though again the network
contains as many super-spreaders and super-receivers as in the original
network. 
Clearly, the absence of in-out correlations makes difficult the spreading of
information. }

\rev{Notice that in the actual underlying static network, though there is  an important fraction of nodes that are simultaneously 
super-spreaders and super-receivers (see diagonal in Fig.~\ref{fig:degree_correlations}B), 
there are also many nodes that are either super-spreader or super-receivers, 
but not both (off-diagonal elements in Fig.~\ref{fig:degree_correlations}B). 
This explain why we observe in Fig.~\ref{fig:firstmoments} that the actual dynamics falls in between 
these two extreme cases, i.e., in between the fully correlated and fully uncorrelated underlying static network.}

If information travels in both direction on edges, we have to make use of 
Eq.~(\ref{eq:pkuncorrelated}). Its associated generating function reads:
\begin{eqnarray}\label{eq:tildeg}
\tilde{g}(x,\tau)&=& \sum_{k'} p_{\infty}(k') \left(1+(x-1)\tilde{T}(\tau)\right)^{k'} \, , 
\end{eqnarray}
where as explained above, $p_{\infty}(k')$ refers to the degree distribution
of the undirected underlying network and $\tilde{T}(\tau)$ to the probability that an
undirected edge is used during $\tau$. 
%
%
To obtain $\tau_c$, we make use of the well-known percolation criterion for uncorrelated undirected networks, 
$\langle k^2 \rangle /\langle k \rangle \geq 2$~\cite{cohen2000resilience}.
As before, if we assume that $\tilde{T}(\tau) \sim \langle \tilde{\rho} \rangle \tau$, then:
\begin{eqnarray}\label{eq:uncorrelatedthreshold}
\tau_c = \frac{\langle k \rangle_{\infty}}{\langle \tilde{\rho} \rangle (\langle k^2
  \rangle_{\infty}-\langle k \rangle_{\infty})}\, .
\end{eqnarray}
This result has been derived by Newman in~\cite{newman2002spread} and recently used in the context of a mobile phone network by Miritello et al. in~\cite{miritello2010dynamical}. 
The estimated value for $\tau_c$ using Eq.~(\ref{eq:uncorrelatedthreshold}) is $12.29$ hours. 
We recall that if the information travels along the direction given by the directed edges, the critical $\tau_c$ corresponds to  
Eq.~(\ref{eq:thresholdcorrelation}). 
The prediction given by Eq.~(\ref{eq:uncorrelatedthreshold})  is close to that obtained from Eq.~(\ref{eq:thresholdfullycorrelated}),  
which corresponds the fully correlated scenario discussed above, see Fig.~\ref{fig:firstmoments}. 
Notice that Eq.~(\ref{eq:thresholdcorrelation}) never reduces to Eq.~(\ref{eq:uncorrelatedthreshold}).  
This indicates that that directed, that is to say, intentional or active information propagation is qualitatively different from unintentional information spreading, i.e., when information travels in both direction along edges. 
For instance, while for unintentional spreading $\tau_c$ depends always on the second moment of the degree distribution, for intentional spreading it may not depend on it, if the network is in-out degree uncorrelated.

\subsubsection*{Size of trees}
In the following we focus on the statistical features of the trees. 
We start out by estimating the size distribution $p(s,\tau)$ under the assumption that information travels in the direction of the (directed) edges. 
In order to get an analytical estimate of $p(s,\tau)$, 
we neglect node-node correlations in the underlying static network as well as
temporal correlations among nodes. 
It will be clear that node-node (topological) correlations can be ignored for
$\tau<\tau_c$, while temporal correlations are always too weak to impact the spreading dynamics (see below).  
We further assume that trees are fully determined by the out-degree
$p(k_o;\tau)$, but as we will see, the assumption breaks down as we approach $\tau_c$. 

These simplifications allow us to estimate the  probability $p(s=1;\tau)$ of
finding a tree of size one as the probability that the
root node has out-degree $0$, i.e., $p(s=1;\tau)=p(k_o=0;\tau)$. The
probability $p(s=2;\tau)$ of finding a tree of size two has
to be equal to the probability that the root node has out-degree $1$ while
simultaneously its unique branch has to lead to a sub-cascade of size $1$, i.e.,
$p(s=2;\tau)=p(k_o=1;\tau).p(s=1;\tau)$. 
More generally, $p(s;\tau)$ is related to $p(s';\tau)$ with $s'<s$.
This relation can be expressed in a compact and elegant way in term of the
generating function  $G(z,\tau)=\sum_{s=1}
p(s,\tau) z^s$ which obeys the following self-consistency equation~\cite{harris2002theory}:
\begin{eqnarray}\label{eq:Gz}
G(z; \tau) = z\, g(1,G(z;\tau);\tau) \,, 
\end{eqnarray}
where $g(1,y;\tau)$ is the generating function of the out-degree distribution
$p(k_o,\tau)$ that is defined as:
\begin{eqnarray} \label{eq:outout}
g(1,y;\tau) = \sum_{k_o} \, p(k_o; \tau)\, y^{k_o} \, .
\end{eqnarray}
%
%
%

%
The cascade size distribution can be obtained from the derivatives of
 $G(z;\tau)$ as 
\begin{eqnarray}\label{eq:Gz_fs}
p(s,\tau) = \frac{1}{n!}\frac{\partial ^{n}G(z;\tau)}{\partial z^{n}}|_{z=0} \,. 
\end{eqnarray}
%
%
In summary, Eq.(\ref{eq:Gz}) provides us  with a method to derive $p(s;\tau)$ 
under the assumption that the  tree
statistics is given by a Galton-Watson (GW) process~\cite{harris2002theory} that
is fully determined by $p(k_o;\tau)$. 
Notice that for a given $\tau$, the out-degree distribution $p(k_o;\tau)$ can be approximated using Eq.~(\ref{eq:generating}) 
as indicated by Eq.~(\ref{eq:outout}). This approximation starts to fail for large values of $\tau$ due to the non-homogenous activity of node over time. 
Alternatively, $p(k_o;\tau)$ can be directly measured from the data for each $\tau$. 

\subsubsection*{Depth of trees}
Now we look for an estimate of the depth distribution $p(d,t)$ under the same assumptions, i.e., 
the tree statistics is given by a GW process fully determined by $p(k_o;\tau)$. 
We define $g_{n+1}(y; \tau) = g_1(g_n(y; \tau); \tau)$, with
$g_1(y;\tau)=g(1,y;\tau)$. 
%
We look for the probability $E_d(\tau)$ that a tree gets extinguished at
depth less or equal than $d$. 
The probability $E_d(\tau)$ obeys $E_d(\tau) = p(0,\tau) +
p(1,\tau)E_{d-1}(\tau) + p(2, \tau) E_{d-1}(\tau)^2 + \hdots + p(k_o,\tau)
E_{d-1}(\tau)^{k_o}+\hdots$. Using a more compact notation~\cite{harris2002theory}, this relation
reads:
\begin{eqnarray} \label{eq:cond_ed}
E_d(\tau) = g_1(E_{d-1}(\tau); \tau)\,.
\end{eqnarray}
On the other hand, the probability that a tree gets extinguished at $d=1$ is directly the
probability that a node does not make phone any phone call in a period $\tau$,
i.e., $E_1(\tau)=p(k_o=0,\tau)$. Thus, using the above given definition of
$g_n(y;\tau)$, we rephrase Eq.(\ref{eq:cond_ed}) as:
%
%
\begin{eqnarray} \label{eq:ed}
E_d(\tau) = g_d(0;\tau) \,.
\end{eqnarray}
We can draw the probability $p(d; \tau)$ from $E_d(\tau)$, as:
\begin{eqnarray}\label{eq:p_d}
p(d; \tau) = E_d(\tau)-E_{d-1}(\tau)= g_d(0;\tau) - g_{d-1}(0;\tau) \,.
\end{eqnarray}

If we assume that information can flow in both direction of the edges, the above depicted GW process for a directed underlying network can be easily adapted to an undirected network, 
the main difference being that the GW process is now fully determined by the (undirected) degree distribution $p(k; \tau)$ --- see Eq.~(\ref{eq:pkuncorrelated}). 
Except for this, the computations of $p(s;\tau)$ and $p(d;\tau)$ follows similar lines, taking into account that a node of degree $k$ can contribute at most with $k-1$ new nodes to the growing tree.

\subsection*{Comparison between theory, original data, and randomized data}

\begin{figure}[!ht]
\centering\resizebox{11 cm}{!}{\rotatebox{0}{\includegraphics{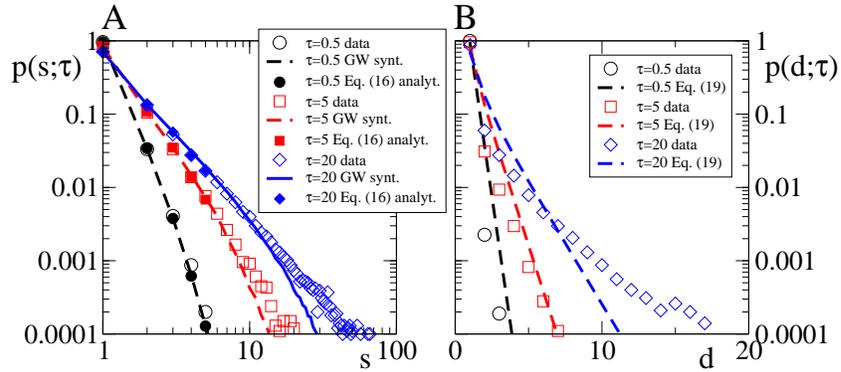}}}
\caption{{\bf Comparison between the tree statistics obtained from
  the transmission theory and real data for short time-scales.}     
The theory provides a reasonable estimate for the size distribution
$p(s;\tau)$, {\bf A}, and depth distribution $p(d;\tau)$, {\bf B}. 
Notice the systematic deviation exhibited by the theory as $\tau \to \tau_C$. 
The discrepancy can be attributed to the absence of topological node-node correlations in the theory. } \label{fig:comparison_dist}
\end{figure}

Fig. \ref{fig:comparison_dist} shows a comparison between Eq.~(\ref{eq:Gz}) (analyt.),
simulations of the proposed GW process (GW synt.), and the tree statistics
obtained from mobile phone data (data). 
The figure indicates that as long as $\tau < \tau_{c}$, the proposed theory
provides a good estimate for the  tree statistics. 
As $\tau \to \tau_c$ (and still for $\tau < \tau_{c}$), the  theory, that neglects (topological) node-node correlations as well as causality effects, 
systematically underestimates the probability of observing large
trees.  
The origin of this discrepancy can be rooted either in the presence of
node-node correlations in the underlying network, or in strong causality
effects arising from temporal correlations. 

\begin{figure}[!ht]
\centering\resizebox{11 cm}{!}{\rotatebox{0}{\includegraphics{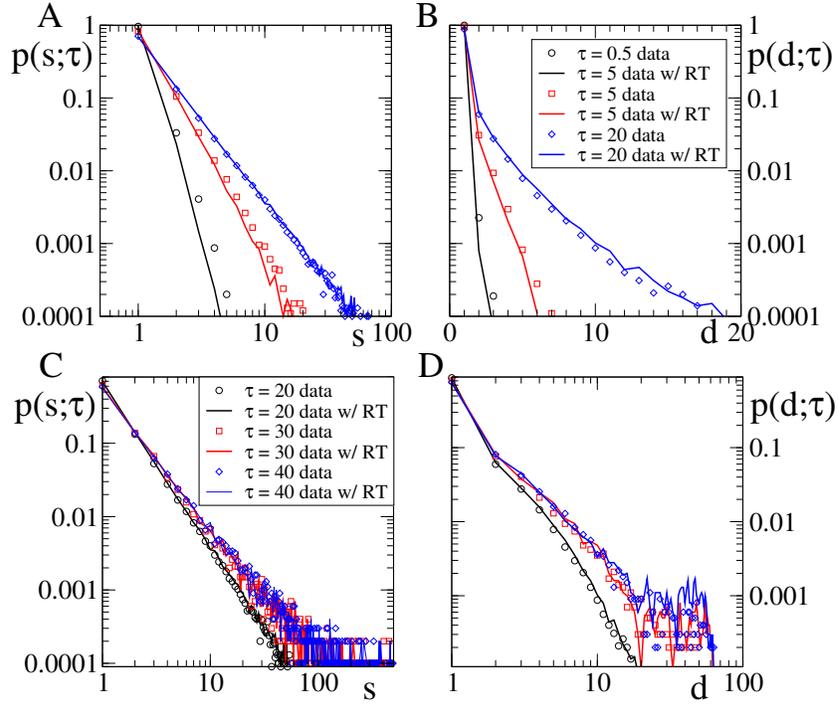}}}
\caption{{\bf Comparison between tree statistics obtained from real data and data with randomized time labels 
   (RT) for small (A and B) and large (C and D) values of $\tau$.} The good matching between real
  and data w/ RT suggests that time correlations do not play a central role on the tree statistics, respectively on 
the spreading process.} \label{fig:largeTau}
\end{figure}

\begin{figure}[!ht]
\centering\resizebox{11 cm}{!}{\rotatebox{0}{\includegraphics{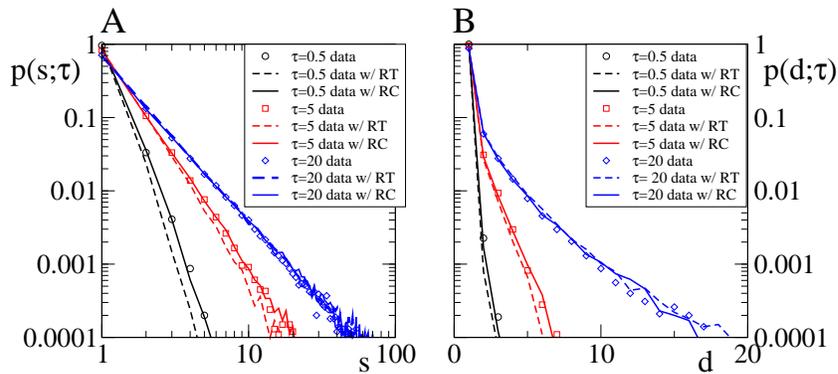}}}
\caption{{\bf Bursty activity of users vs. sender-receiver time-correlations.} 
The difference between real data and data with randomized time labels (RT) is mainly due to the 
absence of bursty activity  in RT data. 
This is evidenced by the good matching between real data and data with random  sender-receiver (temporal) correlations (RC).  
RC data exhibit the same bursty activity of users as the real data. 
} \label{fig:corr}
\end{figure}

\rev{In order to solve this issue, we perform the same tree analysis on two databases that are copies of the original database, where
in one  we have reshuffled the time stamp  phone calls, which we refer to as RT data or data with RT,  
and another where we have reshuffled the order of the phone calls of each user, referred to as RC data or data with RC, see Fig.~\ref{fig:largeTau} and ~\ref{fig:corr}.
More details are provided in the {\it Material and methods} section. 
The use of randomized data  
 is a powerful technique used to define ``null models'' where typically 
some correlations present in the original data are removed by the
randomization procedure. This procedure has been already applied in the context of mobile phone databases~\cite{karsai2010small,miritello2010dynamical}. 
In the RT data, for instance, the topology of the underlying static network is
preserved, as well as the activity edge rate, i.e. $p(\rho)$, 
 day-night and weekly cycles, and in general the global activity patterns,
while the bursty activity of users and sender-receiver temporal correlations are washed out. 
RC data is even closer to the original data, exhibiting even the same bursty activity 
per user, but where sender-receiver temporal correlations are absent.}

\rev{Figs. \ref{fig:comparison_dist} and \ref{fig:largeTau} reveals that the discrepancy
between the proposed theoretical model and the tree statistics obtained from the (original) database, particularly evident for large values of $\tau$,  
 is mainly due to the fact that the theory neglects topological node-node correlations on the underlying static network. 
Trees obtained from the RT data should exhibit statistical features as
the ones predicted by our theoretical  model, except that RT data contains
 node-node (topological) correlations, e.g., degree-degree correlations. 
In other words, any difference between RT data trees and those 
predicted by the theory has to be connected to the presence of node-node
correlations in the RT data. 
The comparison of Figs.~\ref{fig:comparison_dist} and~\ref{fig:largeTau} 
indicates that RT data trees exhibit larger sizes than what the theory predicts as $\tau\to\tau_c$ and above $\tau_c$. 
From this observation we  conclude that topological 
 node-node correlations promote bigger and longer trees. 
As the discrepancy between the theory and RT data increases as we approach  $\tau_c$, 
we learn that these these topological correlations 
  become dominant for large values of $\tau$, particularly above  $\tau_c$. 
On the other hand, for small $\tau$ values, the theory provides a reasonable 
description of the tree statistics. 
This means that at short time scales we can get a rough picture of the tree dynamics 
ignoring topological as well as temporal node-node correlations.} 

\rev{According to what we said above, it is clear that predictions derived from the theory 
are at best equal to the statistics obtained from the RT data. 
On the other hand, the tree statistics obtained from RT data, as shown in Fig.~\ref{fig:largeTau} , 
is remarkably similar to statistics of the original data. 
However, if we observe carefully at very short time scales, the tree statistics corresponding to RT data and original data 
exhibit a small discrepancy, see Fig. \ref{fig:corr}. 
The observed difference cannot be related to topological but to temporal correlations. 
The question here is whether these temporal correlations are due to either the bursty activity of users or to 
sender-receiver temporal  correlations. 
To answer this question we compare the tree statistics computed using RT and RC data, see Fig. \ref{fig:corr}. 
All time correlation have been removed in RT data. On the contrary, RC data exhibit the same bursty activity of users 
present in the original data, while temporal sender-receiver correlations have been washed out. 
Fig. \ref{fig:corr} shows an excellent matching between real and RC data which indicates 
that the discrepancy between RT and real data is due to the absence of bursty activity of user in the former.}  

\rev{In summary, these findings indicate that temporal correlations do not have 
a significant impact on the tree statistics, respectively on the information spreading statistics.  
It is only the temporal correlation coming from the bursty activity of users that affect the tree statistics. 
However, the effect of these temporal correlation is very weak and only observable at very small values of $\tau$. 
In short, temporal sender-receiver correlations seem to play no role. 
Thus, we can safely conclude that  there is no ``large-scale'' causality effects among phone calls, 
i.e., second neighbor correlations seems to be rare, while third neighbor
correlations and so on are virtually nonexistent.  
Does this imply the absence of causality effects? 
Our findings only indicate that causality effects cannot be large-scale and affecting the spreading of information,  
but they may still occur at the local level.}  

\subsection*{Causality loops}
\begin{figure}[!ht]
\centering\resizebox{7cm}{!}{\rotatebox{0}{\includegraphics{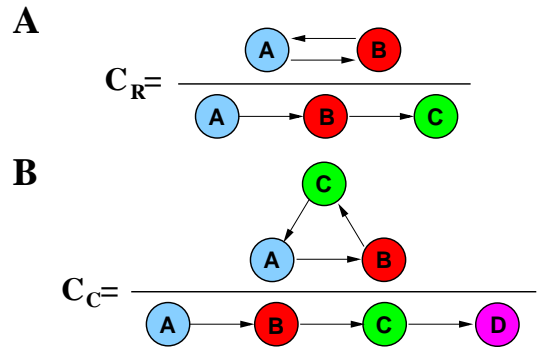}}}
\caption{{\bf Causality loops involve an ordered sequence of phone calls.} The computation of coefficients $C_R(\tau)$ and $C_C(\tau)$ are illustrated in A and B, respectively.} \label{fig:sketch_cycle}
\end{figure}

In the following we explore the possibility of local causality effects in the 
form of causality loops. 
Closed causality patterns do not contribute to the spreading of the information
and are not visible at the level of the tree statistics, since they do not involve the addition of 
new informed users to the set of informed ones.   
We consider two types of patterns:    
the first pattern involves a three-node chain, where user $i$ calls user $j$ at time $t_1$
and user $j$ calls user $k$ at some later time $t_2$, with $t_2 - t_1 \leq \tau$. 
We define the reciprocity coefficient $C_R(\tau)$ as the fraction of three-node chains
where $k=i$, Fig.~\ref{fig:sketch_cycle}A.  
Along similar lines, we define the dynamical clustering coefficient $C_{C}(\tau)$ as the 
fraction of four-node chains where the first and last node are the same,
see Fig.~\ref{fig:sketch_cycle}B.  
%
%

%
%

\begin{figure}[!ht]
\centering\resizebox{11 cm}{!}{\rotatebox{0}{\includegraphics{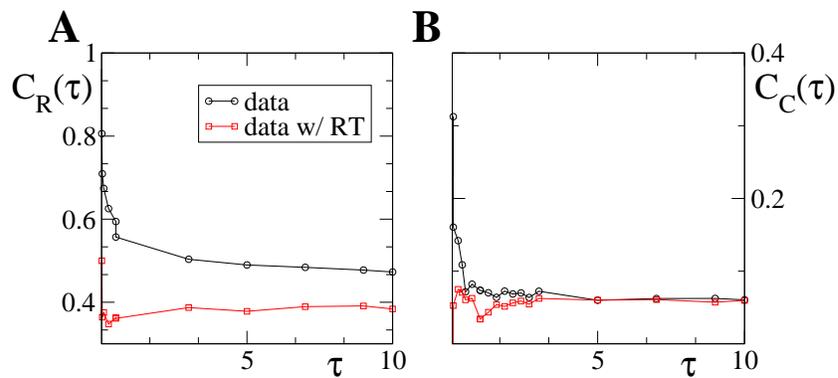}}}
\caption{{\bf Causality loops vs $\tau$.} Reciprocity, {\bf A},  and 
  dynamical clustering, {\bf B}, coefficient for real and time-label randomized (RT)
  data. 
The difference between the curves  is exclusively due to the presence of time
correlations in the real data. 
} \label{fig:clustering}
\end{figure}

Fig.~\ref{fig:clustering} shows that  $C_{R}(\tau)$ and $C_{C}(\tau)$ 
converge for $\tau \to \infty$ to an asymptotic value for both the original and RT
data. 
Though the number of three-node and four-node chains increase monotonically
with $\tau$, the fractions $C_R(\tau)$ and $C_C(\tau)$,  corresponding to
closed 
loops, reach asymptotic values, which indicates that closed chains grow at the same
rate with $\tau$. 
The curves $C_R(\tau)$ and
$C_{C}(\tau)$  
for  RT data corresponds to the fraction of cycles, involving two and three nodes,
respectively, 
expected in absence of causality effects and induced  by the topology of the
underlying static network and edge activity rate $p(\rho)$. 
We observe that at short time scales the values of $C_R(\tau)$ and
$C_{C}(\tau)$ 
obtained from the original data are well above those obtained from RT
data. 
The abundance of causality loops in the original data with respect to RT data, 
reveals that at short time scales the original data exhibits  strong causality 
effects.
Interestingly, the asymptotic  $C_R(\tau)$ value for the original and RT data do not
coincide, being always larger for the original than for RT data.  
This indicates that for any value of $\tau$  the number of reciprocal phone
calls in the original data is larger than what is expected in the absence of
time correlations. 
This finding is likely to be related to the typical message-reply dynamics 
observed for instance in email data~\cite{barabasi2005origin, vazquez2006modeling}. 
On the contrary, the number of three-node loops in the original data seems to converge
asymptotically with $\tau$ to the expected value in the absence of correlations.  
\rev{Finally, we point out that here we have studied the behavior of two particular temporal motifs with $\tau$, but certainly 
more complex motifs are present in the data, see for instance~\cite{kovanen2011}.}

\section*{Discussion}
We have shown that the mobile phone data (as many other communication data) can be represented by a directed (dynamical) network, 
and argued that intentional information spreading requires information to flow in the direction given by the directed edges. 
We have explored this possibility and studied the topological properties of causality trees, such as size and depth, as a proxy to understand information propagation. 
We have introduced a time-scale in the system, the monitoring time $\tau$, which provides a tolerance time 
that allows us to relate two phone calls as causally linked. 
The properties of the causality trees have been studied as function of this time-scale $\tau$. 
Our first observation is that the representation of the data in terms of directed edges reveals 
the existence of super-spreaders and super-receivers.  
We have shown that the tree statistics, respectively the information spreading process, are extremely sensitive to the in-out degree correlation of the users. 
Moreover, we have clearly pointed out that the spreading dynamics under the assumption of intentional spreading is qualitatively different from that obtained under the assumption of unintentional spreading, i.e., when information flows in both direction along edges (see Eqs.~(\ref{eq:thresholdcorrelation}),~(\ref{eq:thresholdfullycorrelated}), ~(\ref{eq:thresholdfullyuncorrelated}), and~(\ref{eq:uncorrelatedthreshold}), and discussion below these equations).
The good matching at short time-scales between the tree statistics  obtained from the original data and the theoretical predictions that neglect time correlations and topological node-node correlations has allowed us to conclude that 
none of these correlations have a strong effect on the tree statistics.
This means that at short time-scales trees can be roughly described as a simple GW process. 
However, at larger time-scales the tree statistics can no longer be explained in terms of this simple theory. 
The tree statistics obtained from randomized time-stamp data indicates that topological node-node correlations, present in the original data but neglected in the theory, dominate the spreading dynamics at these time-scales. 
Moreover, we have learned that these topological correlations promote bigger trees. 
\rev{Time correlations, on the other hand, do not seem to play a significant 
role on the  statistics of tree size and depth at any time-scale. 
Only time correlations coming from the bursty activity of users have an impact on the tree statistics, but this effect 
is very weak and only visible at very short $\tau$ values. 
On the other hand, temporal sender-receiver correlations do not affect the tree statistics.}   
These findings together with the observation that a given information, e.g., a rumor, would require users to 
 retransmit it for more than 30 hours in order to 
cover a macroscopic fraction of the system, suggest 
that there is no intentional broadcasting of information. 
In fact, the  very idea that there is 
information spreading beyond nearest and second-nearest neighbors, i.e., beyond a small vicinity, is called into question.  
At the local level, however, we have observed that  time correlations enhance the number of dynamical closed patterns, an effect particularly evident at short time scales. 
It is only at this level that genuine causality effects, and consequently intentional information propagation, are detectable. 
Nevertheless we stress that these observations apply exclusively to local information circulation.

The analysis performed here can be applied to other communication network data 
like blog and email data~\cite{gruhl2004information,leskovec2006cascading,cointet2009socio,barabasi2005origin,vazquez2006modeling,
malmgren2008poissonian}
, as well as mobile phone data in the presence of exceptional events like natural disasters~\cite{bagrow2011collective}, where different tree structures, and consequently tree statistics, are likely to emerge. 
Finally, the cascade theory we have implemented here applies to  
directed networks in the absence of node-node (topological) correlations.  
Generalizations to account for node-node correlations along the lines of~\cite{boguna2005generalized}  should be possible. 

\section*{Materials and Methods}

\subsubsection*{Mobile phone dataset}
The mobile phone data we have analyzed corresponds to one month of phone calls from a European mobile phone provider. 
To guarantee confidentiality, phone numbers were anonymized. 
This represents 1.044.397 users, that form a connected component, and 13.983.433 phone calls among these users.  
Since our goal has been to study  information transmission, we have constrained 
ourselves to the study of ``successful'' phone calls --- i.e., those where the
receiver has answered the phone call.
There is an average activity of $5.2$ phone calls per second, which leads to an average of $1.7\,10^{-6}$ phone calls per second per directed edge. 
The underlying static network is characterized by $\langle k_i \rangle_{\infty}= \langle k_o \rangle_{\infty} = 2.86$,  $\langle k_i^2 \rangle_{\infty} = 28.4$ and $\langle k_o^2 \rangle_{\infty} = 17.3$. 
Using undirected edges we obtain $\langle k \rangle_{\infty}=4.24$ and $\langle k^2 \rangle_{\infty}=45.05$, and average edge activity $\langle \tilde{\rho} \rangle=2.35\,10^{-6}$ phone calls per second.  
Some aspects of this dataset have been described in~\cite{lambiotte2008geographical},
and some features of the underlying static network has been analyzed in~\cite{stoica2009structure}.

\subsubsection*{Random data -- Null models}

We have performed the data analysis on three datasets: the original data set and two 
 copies of it, one where we have reshuffled the time stamps of the phone calls, which we refer to as RT data, and another where 
we have randomized the order to the phone calls of every user, which we refer to as RC data.  
The RT dataset is an exact copy of the original data set, where source and destination of every phone call remains the same, but the time-stamp of  phone calls are 
randomly exchanged. 
The new dataset is then ordered according to the new time-stamps. 
As result of this procedure, every node exhibits the same in- and out-degree as in the original data set. 
Moreover, the activity rate per (directed) edge and user remain the same, as well as the global activity rate of the dataset (day-night and weekly cycles, etc). 
\rev{The RC dataset is even closer to the original data set. Here we take for every user the vector: (time-stamp, id of callee) generated by the user and reshuffled within this vector the callee id. In other words, we have randomized the order of the phone calls done by each user. 
In the RC dataset users exhibit the same bursty (calling) activity as in the original dataset. }

\section*{Acknowledgments}
We thank E.~Altmann, C.F.~Lee, and F.~Vazquez for valuable comments, and the Max Planck Society for financial support.





\section*{Figure Legends}


\section*{Tables}

\end{document}